\documentclass{esannV2}
\usepackage{graphicx}
\usepackage[latin1]{inputenc}
\usepackage{amssymb,amsmath,array}
\usepackage{subfigure}
\usepackage{url}

\begin{document}
\title{Effect of context in swipe gesture-based continuous authentication on smartphones}

\author{Pekka Siirtola$^{1}$, Jukka Komulainen$^{2}$ and Vili Kellokumpu$^{2,3}$ \thanks{This work was partially funded by the Finnish Foundation for Technology Promotion.}
%
%
\vspace{.3cm}\\
      $^{1}$Biomimetics and Intelligent Systems Group, University of Oulu \\$^{2}$Center for Machine Vision and Signal Analysis, University of Oulu\\$^{3}$Bittium Wireless Ltd\\
\vspace{-\baselineskip}{\tt\small \{pekka.siirtola, jukka.komulainen\}@oulu.fi, vili.kellokumpu@bittium.com}
}

\maketitle


%
%
%
\hyphenation{manu-scripts manu-script ext-re-mums user-in-de-pen-dent user-de-pen-dent in-de-pen-dent mo-del in-de-pen-dent}

\begin{abstract}
This work investigates how context should be taken into account when performing continuous authentication of a smartphone user based on touchscreen and accelerometer readings extracted from swipe gestures. The study is conducted on the publicly available HMOG dataset consisting of 100 study subjects performing pre-defined reading and navigation tasks while sitting and walking. It is shown that context-specific models are needed for different smartphone usage and human activity scenarios to minimize authentication error. Also, the experimental results suggests that utilization of phone movement improves swipe gesture-based verification performance only when the user is moving. 
\end{abstract}

\section{Introduction and related work}
\label{problem}

Traditional "obtrusive" authentication schemes, like passwords, PIN codes and biometrics, do not provide mechanisms to determine whether an active mobile device is being used by the same (or some other) authorized person after the initial access has been granted. Continuous authentication (CA), also referred to as active or implicit authentication, aims at verifying that a device is being used by a legitimate person after login by constantly monitoring the built-in sensor and device usage data, like (partial) face images, touchscreen gestures, device motion, power consumption, in the background (transparently to the user) \cite{patel2016}. 

Touchscreen gesture-based user verification has been a popular approach in CA \cite{patel2016}. Touch input is directly related to the actual physical interaction with the mobile device, thus could be probably used for fast intrusion detection. Most of the existing works have focused on analysing single-finger swipes, i.e. drag and flick, but also other single and multi-finger gestures, like tap/typing, pinch and spread, could be used for touch biometrics. Intuitively, unique phone motion patterns may be observed while user is using touchscreen, thus joint analysis of touch and consequent motion signals has been proposed for CA \cite{kumar2016}. 

A major limitation with prior works on touch biometrics, and CA in general, has been that phone usage and human activity contexts have not been properly taken into account. It can be expected that touchscreen gestures and phone movement patterns have significant differences depending whether the user is browsing or reading (phone usage), or is stationary or moving (human activity), which suggests that CA systems need to be context-aware. Furthermore, phone usage context also defines whether authentication should be performed in the first place. For instance, user verification is not probably needed for casual browsing, while it is crucial if private or confidential data is being accessed \cite{Khan2014}.

The preliminary studies \cite{Khan2014,Mondal2015,Murmuria2015,Feng2015} have demonstrated that application or task specific (phone usage context) modelling can indeed boost the performance of swipe-based CA, while only marginal improvement has been achieved when human activity context has been considered \cite{Feng2015}. So far, human activity context-based models have shown to be useful only when CA is performed based on solely phone movement patterns \cite{lee2017} or combined with typing (tap gestures) \cite{sitova2016hmog}.

In this work, we investigate the role of context when CA is conducted based on touchscreen and accelerometer readings extracted from swipe gestures. Our experimental analysis is performed on the publicly available HMOG dataset \cite{sitova2016hmog} consisting of 100 subjects each performing pre-defined reading and navigation tasks while sitting and walking. We show that both phone usage and human activity context should be considered in swipe gesture-based CA. In addition, our findings suggest that swipe-based CA should rely solely on touch signals when the user is stationary, while inclusion of phone movement patterns improve CA performance only when the user is moving.  

\begin{figure}[h]
\vspace{-1em}
\begin{center}
        \subfigure[Touch feature distribution in read (left) vs. navigation (right) phone usage scenarios.]{\label{f1}\includegraphics[width=0.485\columnwidth]{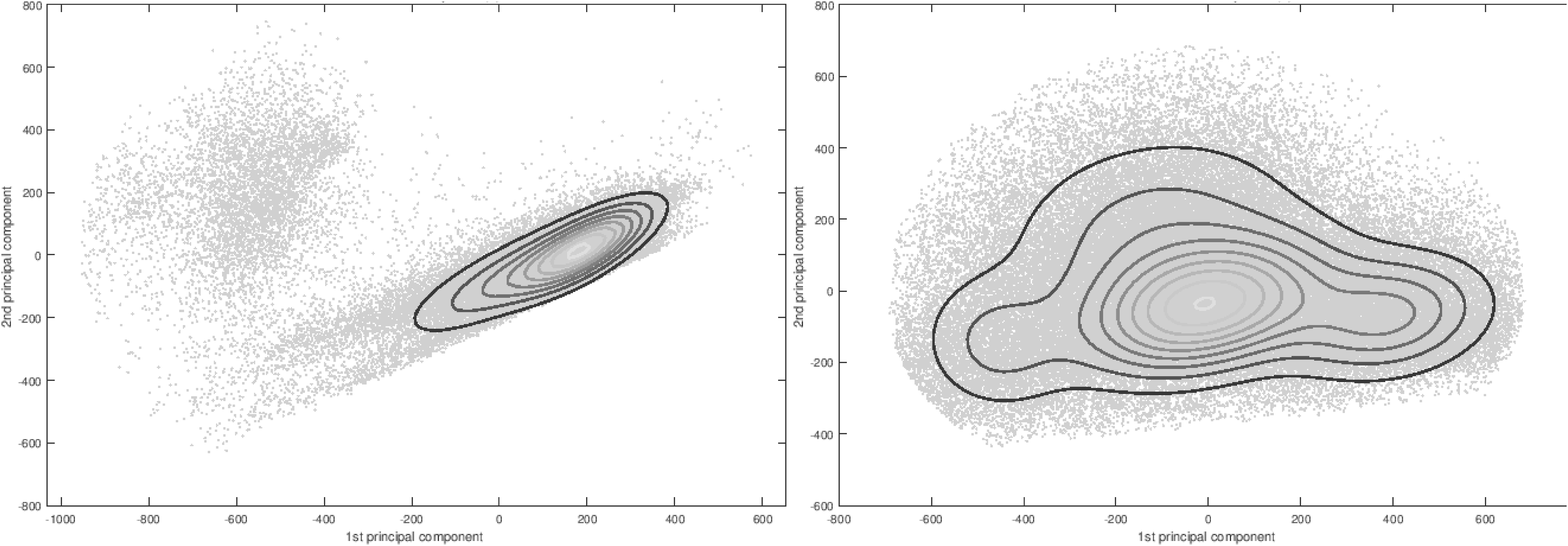}}\hfill        
		\subfigure[Motion feature distribution in two human activities, sitting (left) vs. walking (right).]{\label{f2}\includegraphics[width=0.485\columnwidth]{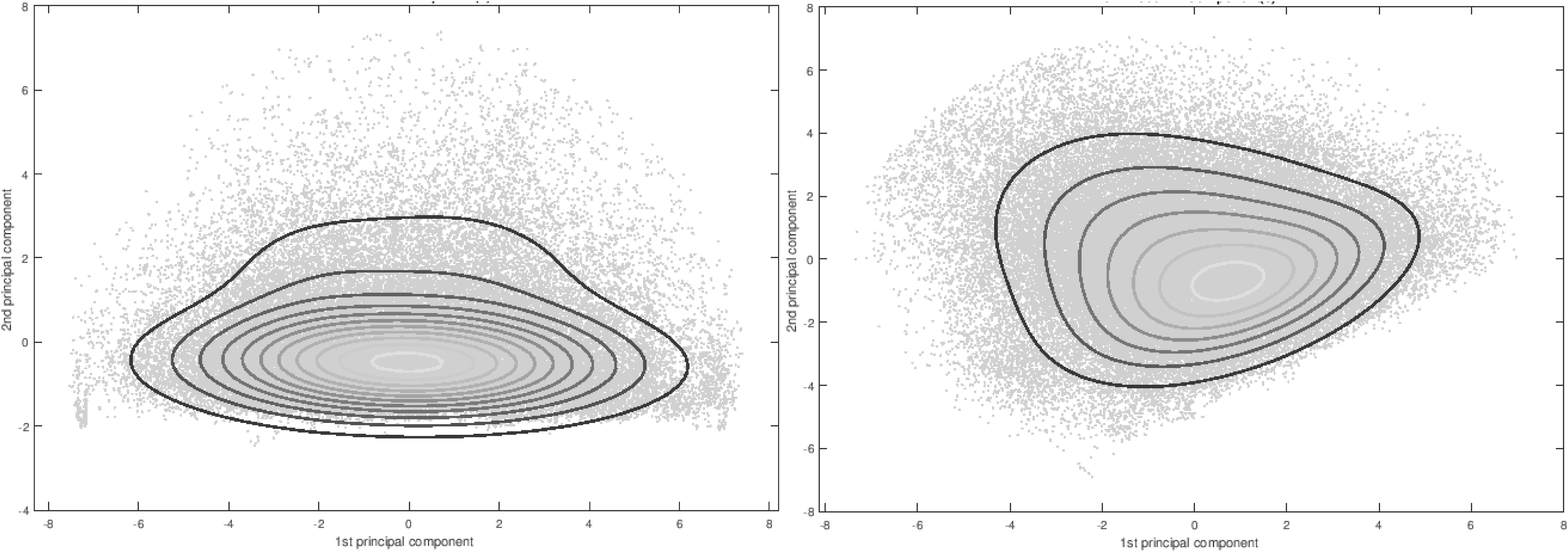}}
        \vspace{-1em}
    \caption{The first two principal components and the corresponding Gaussian mixture models of touch and phone movement based swipe features computed over all subjects in the HMOG dataset \cite{sitova2016hmog}, highlighting the importance context-aware modelling in different a) phone usage and b) human activity scenarios.}
\label{mobileactivity}
\end{center}
\vspace{-2em}
\end{figure}

\section{Methodology}
This study focuses on analysing two different types of context related to: (1) phone usage (reading a document and navigating) and (2) user's physical activity (using the phone while sitting and walking), and their effect in CA of a smartphone user. Figure \ref{mobileactivity} illustrates the distributions of features extracted from touchscreen gestures in the read and navigation scenarios and the features extracted from accelerometer signals during swipe gestures while the phone user is sitting and walking. Based on Figure \ref{f1} it is obvious that touch gestures have indeed significant differences across different phone usage contexts as already shown experimentally in \cite{Khan2014,Mondal2015,Murmuria2015,Feng2015}. Similarly, Figure \ref{f2} depicts that the distribution of accelerometer data features extracted during swipe gestures is highly dependent on the phone user's physical activity. Motivated by this, the aim of this paper is to show that separate models need be trained for each phone usage context and for each type of human activity for accurate CA of a smartphone user.
In the following, we introduce the feature extraction and anomaly detection methods used in our experimental analysis.

\subsection{Feature extraction}
\label{data}

Touchscreen data consists of time stamp, touch pressure and $x$- and $y$-coordinates of a swipe, while accelerometer data contains time stamp, and acceleration signal in $x$, $y$ and $z$ directions. In our study, swipe gestures with more than five touchscreen data points are considered for feature extraction. Altogether, 211 features\footnote{The source codes for feature extraction and classification are available  at: \url{http://www.oulu.fi/bisg/node/40364}} (117 from touchscreen data and 94 from acceleration data) are extracted from these signals. The features extracted from touch coordinates include e.g. mean, variation, percentiles, time, length, velocity, direction, gesture shape related features, start and end point of the swipe in $x$ and $y$ direction as well as from 2D data. Moreover, mean, variation, percentiles, and shape related features are computed from the touch pressure signal. The accelerometer features are calculated from the magnitude signal during swipes and also from 0.5 seconds before and after each swipe, separately. These features include e.g. mean, minimum, maximum, variation, and percentiles. 

\subsection{Anomaly detection}
\label{aim}

We use a simple distance-based one-class ensemble classifier for continuous authentication that is based on the method presented in \cite{smith}. The idea is that swipes locating in the normal region are considered to originate from the genuine user and swipes outside of this are considered as anomalies, and therefore, originating from an impostor. 

The training begins with person-specific feature ranking based on inter-class distance of genuine user's and impostors' data. The features are first normalized to interval 0-1 and the training data $S_{train} = \{ f_1,  \ldots, f_n\}$, where $f_i$ is feature vector $i$, is divided to two sets: data from genuine user $S_{Train_G} = \{ f_{1G},  \ldots, f_{nG}\}$ and data from impostors $S_{Train_I} = \{ f_{1I},  \ldots, f_{nI}\}$. The aim is to find features $f_i$ that differ the most between these two. This is done by calculating mean of each feature separately for both sets after removing outliers. Features can then be ranked by comparing the ratios $(\| mean(f_iG) - mean(f_iI) \|) / mean(f_iG)$. Note that the data from both classes is required in this phase, while the rest of the model training is conducted using only the information from the genuine class.

The actual anomaly detection model is trained using expectation maximization (EM) clustering. The idea in this distance-based classifier is to cluster training data from genuine user's swipe gestures and assume that normal data lies close to the cluster centroid, while anomalies are further away from the centroid. Model used in this study is an ensemble classifier so several classifiers are trained. In this stage, only training data from the genuine user, $S_{Train_G}$, is used and the 40 highest ranked features in the previous stage are selected to train the classifiers. Each classifier is trained using two features: the first classifier uses highest and second highest ranked features, the second uses second and third highest ranked features, and so on. The data from each pair is clustered using EM algorithm, and when a new swipe is classified, its Euclidean distance $d_i$ to each cluster centroids is calculated. The classifiers are combined by calculating the sum $D= \Sigma d_i$ from the distances as suggested in \cite{kittler}.

A threshold for $D$ defines whether the new swipe is classified as normal or an anomaly. This threshold is determined based on training data $S_{Train_G}$. The sum of distances to cluster centroids is calculated to each swipe of $S_{Train_G}$. Let us mark this vector of sum of distances as \textbf{$D_G$} $= \{D_{G1}, \ldots, D_{Gn}\}$, where $n = \| S_{Train_G} \|$. The threshold is then considered as the $i$th percentile of \textbf{$D_G$}. The value of $i$ is optimized separately to each user to find equal error rate (EER). As shown e.g. in \cite{Zhang2015}, the classification on observation as user's typical behaviour or anomaly is not based on one single swipe. Instead, a sequence of 25 sequential observations, and their distance to cluster centroids $C=\{D_1, \ldots, D_{25}\}$, are analysed. In order to avoid unnecessary false alarms, the values of $C$ are ordered and the final classification is based on the mean of four smallest value of $C$.

\section{Experimental setup and analysis}
\label{experiments}


For our study, we considered the HMOG dataset \cite{sitova2016hmog} because it is the only publicly available multi-modal dataset for CA that contains data from both touch gesture and other sensors, including accelerometer, gyroscope and magnetometer. The dataset was collected from 100 subjects who were asked to carry out three different pre-defined tasks using a Samsung Galaxy S4 smartphone: (1) read a document, (2) type text, and (3) use map application. 
Since this study concentrates on swipe gesture-based user authentication, we discarded the data related to the typing scenario. Similarly, touch events related to tapping were not used in the study. Thus, experiments are based on four scenarios: reading a document while sitting (S1), reading while walking (S2), navigating while sitting (S3), and navigating while walking (S4). Each task was performed during four sessions. The swipe gesture data from first two sessions are used for training and the remaining two are used for testing. Experiments are performed using three feature sets: touchscreen, accelerometer and combination of both.

Table \ref{results} reports the average EERs across subjects for context-specific and general CA models. The context-specific models are trained and tested on the data from the same scenario but from different sessions. The general model is trained using data from all scenarios and tested on each individual scenario in turn. According to the results, the results are significantly better when context-specific models are used. In fact, with each scenario, the average EER is lower when the model is trained using the data from same scenario as it is tested compared to the error rate of the general model. Especially the reading scenarios benefit from specific models as the average EER drops from 21.3\% to 11.7\% in S1 and from 15.8\% to 7.0\% in S2 when context-specific authentication model is used instead of the general one.

The results presented in Table \ref{results} suggest that features extracted from acceleration signals are advised to be used only if user is physically active. In fact, it can be seen from Table \ref{results} that the best EER for sitting scenarios are obtained using only touchscreen features while the best EERs for walking scenarios are obtained using fusion of touch and motion features. The results depict also that navigation scenario is more difficult than reading scenario from CA point of view. The main reason for this is that the navigation scenario contains swipes to different directions while in reading scenario swipes are mainly vertical. 
In fact, we conducted an additional experiment with S3 data where horizontal swipes were removed. The EER for touch features dropped from 21.5\% to 17.2\%, which suggests that vertical swipes contain more unique variations among the subjects.


\begin{table}[t]
\caption{EERs (and standard deviations) for context-specific and general models.}
\vspace{-0.5em}
\begin{footnotesize}
\begin{center}
\begin{tabular}{p{2.05cm}p{2.3cm}|p{1.9cm}p{1.9cm}p{1.9cm}}
 \textbf{Train} &  \textbf{Test} &  \textbf{Touch} &  \textbf{Motion} &  \textbf{Fusion}\tabularnewline
 \hline
S1: read \& sit & S1: read \& sit & \textbf{11.7\%} (13.6) & 18.3\% (18.8) & 14.2\% (16.9)\\
S1-S4: all & S1: read \& sit & 25.5\% (19.6)& 29.2\% (21.7)& 21.3\% (17.6)\\
\hline
S2: read \& walk & S2: read \& walk & 9.1\% (12.6) & 11.4\% (12.2) & \textbf{7.0\%} (11.8)\\
S1-S4: all & S2: read \& walk & 26.6\% (18.0)& 21.4\% (15.5) & 15.8\% (12.7)\\
\hline
S3: map \& sit & S3: map \& sit & \textbf{21.5\%} (12.2) & 24.1\% (17.8) & 23.0\% (17.4)\\
S1-S4: all & S3: map \& sit &28.4\% (14.2)& 27.1\% (16.7)& 22.7\% (15.4)\\
\hline
S4: map \& walk & S4: map \& walk & 21.7\% (11.3) & 14.5\% (14.5) & \textbf{13.2\%} (13.4)\\
S1-S4: all & S4: map \& walk &27.4\% (12.5) & 20.8\% (15.7)& 16.8\% (13.5)\\
\hline
\end{tabular}

\label{results}
\end{center}
\end{footnotesize}
\vspace{-2em}
\end{table}

\begin{table}[t]
\caption{EERs (and standard deviations) for cross-scenario tests.}
\vspace{-0.5em}
\begin{footnotesize}
\begin{center}
\begin{tabular}{p{2.05cm}p{2.3cm}|p{1.9cm}p{1.9cm}p{1.9cm}}
 \textbf{Train} &  \textbf{Test} &  \textbf{Touch} &  \textbf{Motion} &  \textbf{Fusion}\tabularnewline
 \hline
S1: read \& sit & S2: read \& walk & 12.2\% (13.6) & 37.2\% (23.3) & 25.2\% (21.8)\\
S2: read \& walk & S1: read \& sit & 11.5\% (13.2)& 40.3\% (19.7)& 26.7\% (22.3)\\
\hline
S3: map \& sit & S4: map \& walk & 25.1\% (12.8) & 34.7\% (19.6) & 32.5\% (20.0) \\
S4: map \& walk & S3: map \& sit & 23.7\% (12.7)& 33.3\% (18.8) & 31.8\% (18.6)\\
\hline
S1: read \& sit & S3: map \& sit & 40.8\% (17.1) & 28.5\% (18.8) & 28.0\% (18.0)\\
S3: map \& sit & S1: read \& sit & 40.7\% (23.7)& 26.2\% (22.8)& 24.4\% (22.4)\\
\hline
S2: read \& walk & S4: map \& walk & 39.1\% (15.0) & 19.2\% (15.0) & 24.4\% (16.6)\\
S4: map \& walk & S2: read \& walk & 40.8\% (23.4) & 22.1\% (19.7)& 20.9\% (18.7)\\
\hline
\end{tabular}

\label{results2}
\end{center}
\end{footnotesize}
\vspace{-2em}
\end{table}

Table \ref{results2} demonstrates the importance of considering both phone usage and physical activity contexts in the training process. 
The results correspond to the average EERs when CA models are trained with one phone usage context and tested with the other while the physical activity remains the same. In addition, Table \ref{results2} presents the average EERs when CA models are trained with one physical activity and tested with the other while the phone usage context remains the same. The result comparison between Table \ref{results} and Table \ref{results2} shows as well that CA models should be context-specific. For instance, the phone usage context has a huge effect on the error rates as the EER for S2 jumps from 7.0\% to 20.9\% when training is performed on navigation while walking scenario (S4) instead of reading while walking (S2). Similarly, the EER for S2 increases from 7.0\% to 12.2\% when training is performed on reading while sitting scenario (S1) instead of reading while walking (S2). In general, it can be noted that the phone usage context has more significant effect on error rates than user's physical activity.


\section{Conclusions and future work}
\label{conclusion}
This study investigated the role of context in swipe gesture-based continuous authentication. It was shown that both phone usage and human activity scenarios affect how the user is interacting with the device. Therefore, not only different applications or tasks require their own specific model but different human activities as well. According to the experiments, only touch features should be used for swipe-based verification if the user is stationary and combination of features extracted from touchscreen and accelerometer signals when the user is moving.

We plan to extend our work from laboratory conditions into unconstrained real-world scenarios where context information is not available. In general, the phone usage context can be determined based on the currently running foreground application, while we believe that the human activity context can be estimated using pre-trained models \cite{ARready} or unsupervised learning methods.


\begin{footnotesize}

\end{footnotesize}

\end{document}